\documentclass[conference]{IEEEtran}
\IEEEoverridecommandlockouts
\usepackage{cite}
\usepackage{amsmath,amssymb,amsfonts}
\usepackage{algorithmic}
\usepackage{graphicx}
\usepackage{textcomp}
\usepackage{xcolor}
\graphicspath{{figuresGM/}}
\usepackage{url}

\def\BibTeX{{\rm B\kern-.05em{\sc i\kern-.025em b}\kern-.08em
    T\kern-.1667em\lower.7ex\hbox{E}\kern-.125emX}}
\begin{document}

\title{A Two-Layer Framework with Battery Temperature Optimal Control and Network Optimal Power Flow\\

\author{Anshuman Singh, Wang Peng, and Hung D. Nguyen$^\star$
\thanks{$^\star$ Corresponding author}
\thanks{Authors are with School of EEE, NTU, Singapore. \textit{anshuman004@e.ntu.edu.sg, epwang, hunghtd@ntu.edu.sg}}}
}



\maketitle

\begin{abstract}
Battery energy storage is an essential component of a microgrid. The working temperature of the battery is an important factor as a high-temperature condition generally increases losses, reduces useful life, and can even lead to fire hazards. Hence, it is indispensable to regulate the temperature profile of the battery modules/packs properly in the battery energy storage during the operation. In view of this, a two-layer optimal control and operation scheme is proposed for a microgrid with energy storage. In the first layer, an optimal control model is formed to derive the optimal control policy that minimizes the control efforts, consisting of the fan speed and battery current magnitude, in order to achieve a temperature distribution reference over the battery modules. In the second layer, the system operator of the microgrid performs an optimal power flow to search for the optimal temperature distribution reference used in the first stage and the corresponding operating current of the battery that minimize the operation cost of the entire microgrid system. This two-layer scheme offers a great computational benefit that allows for large-scale integration of batteries. A case study is performed on the proposed two-layer model to illustrate its performance.
\end{abstract}



\section{Introduction}
The presence of battery energy storage in a microgrid increases the flexibility of operation. With the global trend of higher integration of renewable energy sources that are variable and intermittent, the role of the energy storage system (ESS) has become more crucial \cite{barton2004energy}. The temperature of the individual cells of the battery decides the operating efficiency and the degradation rate. High operating temperatures lead to increased resistance of cells that causes more heat generation inside the cell. Also, the degradation rate of a cell is increased at higher operating temperatures. Hence, the ESS should be operated in conditions where the temperature is in a safe range. 

In general, the optimal power flow (OPF) models for a microgrid having energy storage have only State of charge constraints in their mathematical model \cite{PV_with_Batteries,DC_microgrid,OPF_in_MG,nguyen2012new}. The temperature constraints of battery are usually ignored due to difficulty in the thermal modelling of the individual battery modules and including them in the OPF model. Authors in \cite{Optimal_control_rw} have tried to design a controller for the energy storage system (ESS) of a residential consumer such that the bill of energy consumption is reduced. Here, modelling of battery temperature and aging of battery is also included. Authors in \cite{hector} have tried to formulate an optimal control problem such that the time required to charge the battery and the degradation in that time interval is minimized. Since both the objectives are conflicting, hence weight in the objective decides the control action. In both the above papers, only one battery is considered and implementation at the grid level is difficult because of the large number of parameters involved.

At the level of the battery, thermal modelling is done using simulation software that uses computational fluid dynamics (CFD) modelling. This process is time-consuming and can not be synchronised with the OPF which takes place in shorter time intervals. The problem can be simplified by considering the collection of all battery modules as a single battery and represent it with a single temperature. However, it may lead to underestimation or overestimation based on the model. In this work, a lumped thermal model is presented which analyses temperature at discrete points instead of modelling temperature of each cell inside a battery module \cite{jasinski}.  

The heat is generated in the battery due to resistive heating, over-potential loss, and change in entropy \cite{rosewater}. Since the size of ESS is large at the scale of the grid, only resistive heating is considered as the cause of heat generation. In heat generation due to resistive heating, the amount of heat generated by a battery is directly proportional to the square of the magnitude of the current. Hence, the ESS controller is only concerned about the magnitude of the current flowing through the battery. Unlike other control models that decide when and how much the ESS should charge or discharge, the proposed optimal control model in this paper only optimizes the magnitude of the current flowing through the battery. With this simplification, the explicit optimal control laws can be derived and conveniently used in the network optimal power flow. Then the charging or discharging decision will be made by the system operator at the level of the microgrid.  

The contributions of this paper are summarised as follows.
\begin{itemize}
    \item The dependence of the battery module temperature profile on the battery current magnitude and fan speed is established.
    \item A two-level model having optimal control of batteries and optimal power flow for a microgrid is modelled. This two-level optimization framework can be solved effectively and thus paying the way to incorporate more batteries with complicated and dense mesh for temperature computation in the network's optimal operation. 
\end{itemize}

The remainder of this paper is organised as the followings. In Section \ref{sec:mathform}, thermal modelling of battery modules and network modelling of microgrid is formulated. In Sections \ref{sec:mixopt} and \ref{sec:twolayer}, operation of microgrid using combined formulation and bi-level formulation is presented. A case study is performed on the proposed model and the results are discussed in Section \ref{sec:numsil} where the impact of proposed model on the temperature of the battery and overall cost is measured and compared with the combined formulation. Finally, Section \ref{sec:concl} provides the summary and discusses the future scope of this work.

\section{Mathematical Formulation} \label{sec:mathform}
\subsection{Thermal modelling of battery modules}

\begin{figure}[tb]
\centerline{\includegraphics[width = \linewidth]{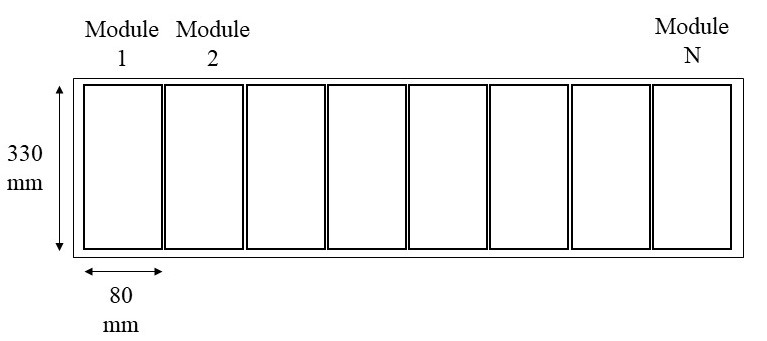}}
\caption{Configuration of N battery modules in a row}
\vskip -1em
\label{fig:config}
\end{figure}

Steady-state modelling of heat transfer is considered in this paper. Figure \ref{fig:config} presents a configuration of battery modules in a battery rack. Each battery module is represented by a node located at the centre of the module and the surface temperature of the battery. It is considered that $N$ battery modules are kept in a straight line without any air gap between them. The energy balance at each node can be written as   
\begin{equation}
    \dot{Q} + \sum_{out} Q = 0\\
\end{equation}
where $\dot{Q}$ is the heat generated inside the battery module. Each node transfers heat to the neighbouring node through conduction from the front and back face. Heat is also lost to air through convection from the side and top faces. The magnitude of heat loss through radiation is small compared to conduction and convection hence ignored to keep equations linear. At steady state, heat transfer through conduction between node $i$ and node $i+1$ can be written as shown in \eqref{cond}. Heat lost to air can be written as shown in \eqref{convection}.
\begin{align}
    Q_{cond} & = \frac{k_b}{l} A_{cond} (T_i - T_{i+1})  \label{cond}\\
    Q_{conv} & = h A_{conv} (T_i - T_a)   \label{convection}
\end{align}
where $T_a$ is the ambient temperature, $l$ is the length of battery, and A is the area perpendicular to the direction of heat transfer. $k_b$ is the conduction heat transfer coefficient of battery casing. $h$ is the convective heat transfer coefficient and assumed to be varying linearly with fan speed $u_f$ as
\begin{align}
    h = h_0 (1 + \lambda (u_f - u_{f_0})).
\end{align}
Heat generation due to resistive heating is only considered in this paper and can be written as
\begin{align}
    \dot{Q} = I_b^2 R_b
\end{align}
where $R_b$ is the resistance and $I_b$ is the current through the battery. 
The top covering of the battery module is considered to be an insulating surface and there is no heat transfer between the bottom surface and the rack. The energy balance for the 1st battery can be written as for $k_b'={k_b A_1}/{l}$
\begin{align}
    \dot{Q_1} = h A_1 (T_1 - T_a) + 2 h A_2 (T_1 - T_a) + k_b' (T_1 - T_2)
\end{align}
The energy balance for all batteries from 2 to $N-1$ can be written as
\begin{align}
    \dot{Q_i} = k_b' (T_i - T_{i-1}) + k_b' (T_i - T_{i+1}) + 2 h A_2 (T_i - T_a) 
\end{align}
The energy balance for the last battery can be written as
\begin{align}
    \dot{Q_N} = k_b' (T_N - T_{N-1}) + h A_1(T_N - T_a) + 2 h A_2 (T_N - T_a)
\end{align}
The equation for node $i$ can be generalised as a linear algebraic equation as shown below
\begin{align}
    b_{i1} T_1 + b_{i2} T_2 + . . . + b_{in} T_N + b_{i (N+1)} T_a = I^2_i R_{b_i}
\end{align}
Collectively, this can be represented by a system of equations 
\begin{align} \label{eq:BTC}
    B \, T = C.
\end{align}
Here, the size of B matrix is $(N,\,1)$ column vector representing the heat generation within N battery modules. 

In the following, we further represent the two coefficients matrices $B$ and $C$ in terms of the control inputs $u = [u_f, u_I]$, which consists of the fan speed $u_f$ and squared current $u_I = I_b^2$, for regulating the temperature $T$ of the battery's cells. The resulting representation will be leveraged to derive the optimal control policy in Section \ref{sec:opt_contrl} that minimizes the control efforts while achieving the desired temperature profile.

The sensitivity matrix B is a linear function of heat transfer coefficient $h$ which in turn is a linear function of fan speed $u_f$. Hence, matrix B can be written as 
\begin{equation}
    B = B_0 + K u_f.
\end{equation} 
The constant matrix $B_0$ is used in the absence of a fan. The coefficient $K$ reflects the effectiveness of the forced ventilation while varying the fan speed $u_f$. 

The matrix C depends on the squared battery current and resistance of battery. The resistance of battery is an affine function of temperature, i.e., $R = R_{ref} \left(1 + \alpha_T (T - T_{ref})\right)$ where the subscript \textit{ref} denotes the reference values, typically at $25 ^{\circ}$. $\alpha_T$ is the temperature coefficient of resistance; for the battery material, we choose $\alpha_T = 0.004$. The resistance of battery can be reduced to $R = R_0 + \alpha T$ where $R_0 = R_{ref} (1 - \alpha_T T_{ref})$ and $\alpha = \alpha_T R_{ref}$. Hence, matrix C can be written as
\begin{align}
    C = u_I (R_0 + \alpha T).
\end{align}

As a result, the temperature profile relation \eqref{eq:BTC} can be rewritten as:
\begin{equation} \label{eq:BTCx}
 (B_0 + K u_f) T = u_I (R_0 + \alpha T).
\end{equation}
Rearranging \eqref{eq:BTCx} and introducing $M = B_0 + K u_f - \alpha u_I$ and $\beta = u_I R_0$, yields:
\begin{equation} \label{eq:BTCM}
 M T = \beta
\end{equation}
which will be used to design the optimal control policy in Section \ref{sec:opt_contrl} below.

\subsection{Network modelling of microgrid}
The relationship between power and voltage is given by the linear branch flow model for all $(i,j)\in U$, where $U$ is the set of all lines.
\begin{align}
P_{ij} + p^g_j-p^d_j = \sum_{k:(j,k)\in U} P_{jk}, \label{OPF1}\\
Q_{ij} + q^g_j-q^d_j = \sum_{k:(j,k)\in U}Q_{jk}, \\
v_j = v_i - 2(r_{ij}P_{ij} + x_{ij}Q_{ij}),  \\
P_{grid} = P_{12}.
\end{align}
The notations are defined below.\\
$p^g_j/q^g_j :$ active/reactive power generated at bus \textit{j} \\
$p^d_j/q^d_j :$ active/reactive power demand at bus \textit{j}, \\
$r_{ij}/x_{ij} :$ resistance/reactance of line between bus \textit{i} and \textit{j}\\
$v_i :$ square of voltage magnitude at bus \textit{i}\\
$P_{ij}/Q_{ij} :$ active/reactive power line flow from bus \textit{i} to  bus \textit{j} 
The constraints related to battery storage are as follows
\begin{align}
&E_{t+1} =  E_t + (P_{bc} \eta_{bc} - P_{bd}/\eta_{bd}) \Delta t,\\
&P_{bc} + P_{bd} =  V_{ess} I_{ess}, \\
&\alpha_{bc} + \alpha_{bd} \leq 1. \label{OPF2}
\end{align}
Here $E_{t+1}/E_t$ is the battery capacity at time interval $t+1/t$ respectively. $\eta_{bc}$ is the efficiency of battery charging and $\eta_{bd}$ is the efficiency of battery discharging. $V_{ess}$ and $I_{ess}$ are overall voltage and current through ESS depending upon series-parallel configuration of battery modules. $\alpha_{bc}/ \alpha_{bd}$ are binary variables associated with charging/discharging process. 

\section{Mixed Optimal operation of microgrid} \label{sec:mixopt}
\subsection{Mixed Optimal Formulation}
Optimal power flow is a fundamental problem for the operation that optimizes the operation cost while satisfying network and operational constraints \cite{nguyen2018constructing, HungGPOPF}. The OPF framework proposed in this work is an extension of the conventional OPF problem that accommodates the battery temperature constraints. This new framework is called the mixed or combined optimal operation formulation because this is a unified optimization formulation crossing the thermal and electrical domains. This framework can be modelled as
\begin{align}
\min \,  C_{grid} + C_{ess}.
\end{align}
$C_{grid}$ is the energy import cost from the utility grid and $C_{ess}$ is the cost of operation of ESS.
\begin{align}
    C_{grid} = \sum_T P_{grid} f_{e} \Delta t\\
    C_{ess} = \sum_T P_{ess} f_{ess} \Delta t
\end{align}
where $P_{grid}$ is the net power exchange from the grid and $f_e$ is the rate of power exchange (\$/MW) from the grid. $P_{ess}$ is the net power exchange by ESS and $f_{ess}$ is the corresponding rate of power exchange (\$/MW). The time window $\Delta t$ is the duration that one shot of the mixed OPF concerns. Typical time window can be 5 minutes. The constraints in this mixed model are network constraints (\ref{OPF1} - \ref{OPF2}), thermal modelling constraints and operational limits.

\subsection{Limitations of the mixed formulation}
The mixed formulation minimizes the cost of operation, mostly from the electrical domain, while satisfying the constraints including constraints on the temperature of each battery module in the thermal domain. This unified optimization problem considers all possible information from both domains and thus bringing potentially better optimal solutions if the optimization can be solved properly. However, there are certain limitations to the implementation of this formulation. Firstly, for a large number of batteries in the network, the size of the optimization model will be huge. If a dense mesh is used for the temperature profile computation, the number of considered temperature points/variables is large. Solving such large-scale and complex optimization problems is not numerically efficient while there is no guarantee that a high-quality optimal solution can be found. Secondly, the system operator may not be aware of the private information within the battery such as the relationships among battery current, fan speed, and temperature distribution. Therefore, this work proposes a two-layer optimization framework to handle the operation in both thermal and electrical domains.

\section{Two-layer Optimization} \label{sec:twolayer}
\subsection{Optimal control of temperature in the thermal domain} \label{sec:opt_contrl}
Assume that the battery's current temperature profile is $T$ and the desired profile is $T^\star$ satisfying the thermal management requirement. To drive the battery's temperature profile to $T^\star$, one needs to control the fan speed as well as the battery's current. Let such required control efforts be $\Delta u = [\Delta u_f, \Delta u_I]$. We introduce the following relation with its proof presented in Appendix \ref{sec:app} 
\begin{equation} \label{eq:eff_rel}
KT \, \Delta u_f - (\alpha T + R_0) \Delta u_I = -M(T^\star - T).
\end{equation}

Introducing $a = -M(T^\star - T)/(KT)$ and $b = (\alpha T + R_0)/(KT)$, \eqref{eq:eff_rel} leads to
\begin{equation} \label{eq:eff_relc}
\Delta u_f = a + b \,\Delta u_I.  
\end{equation}

The optimal control is designed to optimize the control effort $\Delta u = [\Delta u_f, \Delta u_I]$ that sufficiently brings the battery's current temperature profile $T$ to the desired profile $T^\star$.

\begin{equation}\label{eq:opt_contrl}
\begin{aligned} 
 \min_{\Delta u} \quad & (\Delta u_f)^2 + c\, (\Delta u_I)^2  \\
 \text{subject to} \quad & \Delta u_f = a + b \,\Delta u_I \\
\end{aligned}
\end{equation}
where $c>0$ is the weight of the corresponding objective component. This weight can also be chosen to reflect how a change in the fan speed leads to a change in the battery current in case the fan is powered by the battery. 

\subsubsection{Optimal temperature control policy}
The optimal policy that optimizes \eqref{eq:opt_contrl} is as below.
\begin{equation}\label{eq:opt_sol}
   \begin{aligned} 
    \Delta u_f & = a - ab^2/(b^2 + c), \\
    \Delta u_I & = - ab/(b^2 + c).
\end{aligned} 
\end{equation}

To prove this optimal policy, one can replace $\Delta u_f$ in the objective function of \eqref{eq:opt_contrl} by $a + b \,\Delta u_I$ to make it a univariate function of $\Delta u_I$. Taking the derivative of the resulting univariate function with respect to $\Delta u_I$ and setting it to zero, one can arrive at the optimal policy \eqref{eq:opt_sol}. 

Recall that, for a given current temperature profile $T$, the term $a$ depends affinely on the temperature setting $T^\star$. The optimal policy \eqref{eq:opt_contrl} provides an explicit form in terms of the temperature setting $T^\star$, i.e., $\Delta u = \Delta u(T^\star)$. In other words, for a given temperature setting $T^\star$, the optimal policy \eqref{eq:opt_sol} will tell the optimal amount of control effort needed to achieve such a desired temperature profile from the current temperature. 

\subsection{Optimal power flow in the electrical domain}
Once the optimal control policy is established in the lower layer, the optimal control efforts and the temperature profile reference will be further optimized in the upper optimization layer with OPF. The output of the upper layer contains the charging/discharging decision and the temperature reference settings. For the upper layer, the optimization objective is to minimize the total cost of operation.
\begin{equation}
    \begin{aligned}
\min \quad &  C_{grid} + C_{ess} \\
\text{s.t.} \quad & \text{OPF constraints:}\, (\ref{OPF1}) - (\ref{OPF2})\\
     & \text{Opt. control policy}\,\eqref{eq:opt_sol}:\, \Delta u = \Delta u(T^\star)
\end{aligned}
\end{equation}
Here vector $\Delta u$ contains the fan speed and battery current magnitude increments needed to provided at the current state to achieve the desired temperature profile $T^\star$. The operational limit on all system variables can be modelled as an inequality constraint given as
\begin{align}
\underline x_p \leq x_p \leq \overline x_p  \label{xp}
\end{align}
here, $x_p$ is the set of all system variables.

\section{Numerical simulations} \label{sec:numsil}

\subsection{Test system}
\begin{figure}
    \centering
    \includegraphics[width=\linewidth]{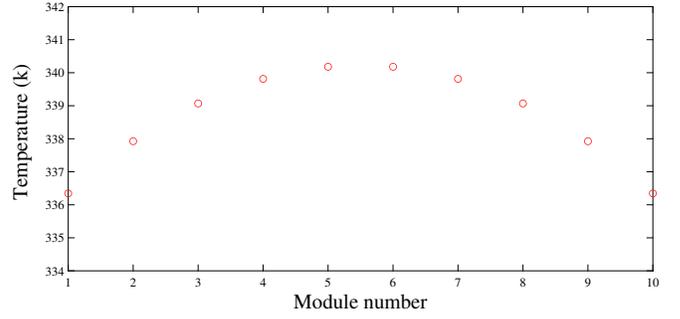}
    \caption{Initial temperature profile}
    \vskip -1em
    \label{T_profile}
\end{figure}

A case study is performed to evaluate the effect of operating battery with and without local temperature control. A 33 bus radial power distribution network is considered \cite{baran1989}. Bus 1 is connected to the utility grid and a battery energy storage of rating 66.304 kWh is considered to be located at bus 6.
There are 10 battery modules of rating 25.9 V/64 Ah each kept in a row constituting one level of the battery rack as shown in Fig. \ref{fig:config}. All the modules in a row are connected in series. There are 4 levels or rows in the rack having a similar configuration and are connected in parallel to each other. Assuming symmetry in the operation of the 4 levels, only the temperature distribution of one row is considered and rest of the rows follow same pattern. Initial temperature distribution is shown in Fig. \ref{T_profile}. The height of each battery module is 230 mm. The other parameters of the test system are given in Table \ref{testdata}. The simulations are performed on MALAB R2019a using Yalmip toolbox and Gurobi as the solver.

\begin{table}[tbp]
    \centering
    \caption{Data of test system}
    \begin{tabular}{c|c|c}
    \hline
        Ambient temperature ($T_a$) & K & 308\\
        Battery voltage rating ($V_{b}$) & V & 259 \\
        Buying/selling rate at grid ($P_{grid}$) & \$/MW & 30/26 \\
        Buying/selling rate for ESS ($P_{ess}$) & \$/MW & 26/32 \\
        Charging efficiency ($\eta_c$) &  \% & 0.95\\
        Coefficient of conduction  $(k_b)$ & W/mk & 205\\
        Coefficient of convection loss ($h_{0}$) & $W/m^2K$ & 5\\
        Initial SOC ($E_t$) & kWh& 40 \\
        Discharging efficiency ($\eta_d$) & \%  & 0.95\\
        ESS rated voltage  & V & 259\\
        Fan speed coefficient of h $(\lambda)$ & min/rotations &  0.01814 \\
        Initial fan speed $(u_f)$ & rpm & 2000\\
        Initial current through each battery ($u_i$) & A & 50\\
        Maximum SOC ($\overline{E}$) & kWh  & 66.304\\
        Maximum charge power of ESS $(\overline{P_{bc}})$ & kW & 60\\
        Maximum discharge power of ESS $(\overline{P_{bd}})$ & kW & 60\\
        Maximum bus voltage ($\overline{V}$) & p.u. & 1.1 \\
        Minimum SOC ($\underline{E}$) & kWh & 5\\
        Minimum bus voltage ($\underline{V}$) & p.u. & 0.9 \\
        Resistance of each module ($R_{ref}$) & ohm & 0.1\\
        Temperature coefficient of resistance $(\alpha_T)$ & $K^{-1}$ &  0.004\\
        \hline
    \end{tabular}
    \label{testdata}
\end{table}

\subsection{Discussion}
\begin{figure}[tb]
    \centering
    \includegraphics[width=\linewidth]{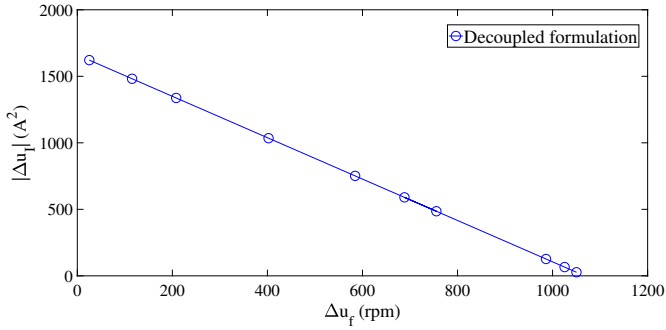}
    \vskip -1em
    \caption{Change in control variables with variation in weight ($T^\star$ = 0.95 T)}
    \label{plot1}
\end{figure}

\begin{figure}[tb]
    \centering
    \includegraphics[width=\linewidth]{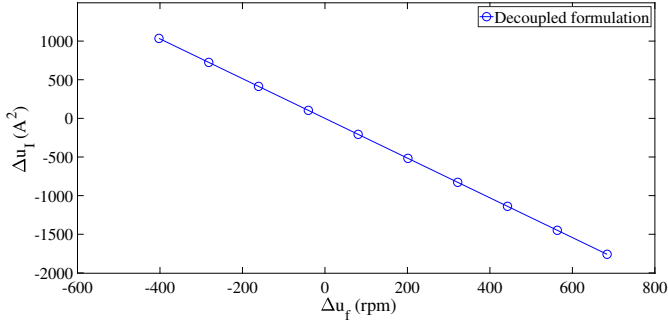}
    \vskip -1em
    \caption{Changes in control variables with variation in desired temperature (for c = 0.25)}
    \label{plot2}
\end{figure}

Fig. \ref{plot1} shows the change in control variables for different values of the weight of objective ($c$) and a given desired temperature of $0.95 T$. The decoupled formulation refers to the two-layer optimization framework. Fig. \ref{plot2} shows the change in control variables for different values of desired temperature for a given weight of 0.25. These figures shows the trade-off between the two control efforts i.e. varying the fan speed versus varying the current. The analysis of control effort is significant because there is a cost associated with each control effort. Operating battery at higher temperatures will provide more output and in turn more revenue but also increases the degradation rate thereby reducing the remaining useful life.

\begin{figure}[tb]
    \centering
    \includegraphics[width=\linewidth]{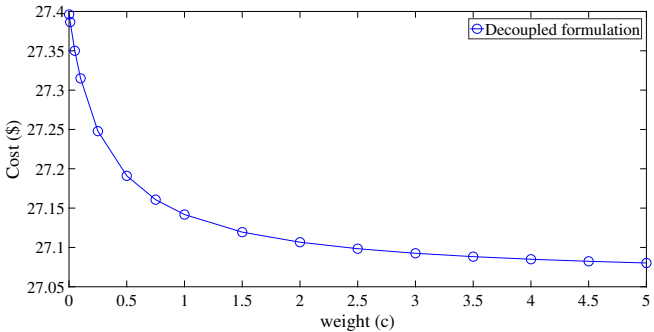}
    \vskip -1em
    \caption{Total cost of operation versus weight (for $T^\star$ = 0.95 T)}
    \label{plot3}
\end{figure}

Fig. \ref{plot3} shows the total cost of operation of the microgrid for different values of the weight of objective ($c$). It is observed that there is not a significant variation in cost. Also, the cost is higher in case of less desired changes in current magnitude. The effects will be more visible when there are more than one ESS and each ESS operator try to locally optimize its fan speed and ESS current.


\section{Conclusion} \label{sec:concl}
In this paper, a simple two-layer mathematical formulation is discussed for optimal operation in a microgrid having energy storage. In the lower layer or layer 1, the decision on values of battery current magnitude and fan speed is taken such that the battery modules' temperature profile reaches the desired profile that is compliant with the thermal management requirement. These values are passed on to the microgrid operator who performs optimal power flow in the upper level, level 2. A case study is performed on the proposed method of operation and the results are discussed. The future scope of this paper includes more detailed thermal modelling of individual battery cells and dynamic thermal modelling of batteries. Also, the modelling of the remaining useful life and the cost associated with the degradation of battery can be included in the optimal control model.

\section{Appendix} \label{sec:app}
This appendix shows the proof of the relation \eqref{eq:eff_rel}. Perturbing \eqref{eq:BTCM}, we have

\begin{equation} \label{eq:pertur}
    (M + \Delta M) (T + \Delta T) = \beta + \Delta \beta.
\end{equation}
Expanding \eqref{eq:pertur} and assuming that $\Delta M \, \Delta T$ is small enough to be ignored, one arrives at the following linearized form:
\begin{equation} \label{eq:lin}
    M\,\Delta T +  T \, \Delta M = \Delta \beta.
\end{equation}
Using $\Delta T = T^\star -T$, $\Delta M = K \Delta u_f - \alpha \Delta u_I$, and $\Delta \beta = R_0 \Delta u_I$, \eqref{eq:lin} becomes
\begin{equation} \label{eq:lin_f}
    M\,(T^\star -T) + (K \Delta u_f - \alpha \Delta u_I) \,T = R_0 \Delta u_I
\end{equation}
which directly leads to \eqref{eq:eff_rel}, i.e.,
\begin{equation*}
    KT \, \Delta u_f - (\alpha T + R_0) \Delta u_I = -M(T^\star - T).
\end{equation*}

\section*{Acknowledgment}
This work is supported by NTU SUG, MOE Tier-1 2019-T1-001-119, EMA \& NRF EMA-EP004-EKJGC-0003.
\bibliographystyle{ieeetr}
\bibliography{GenMeet.bbl} 

\begin{thebibliography}{10}

\bibitem{barton2004energy}
J.~P. Barton and D.~G. Infield, ``Energy storage and its use with intermittent
  renewable energy,'' {\em IEEE transactions on energy conversion}, vol.~19,
  no.~2, pp.~441--448, 2004.

\bibitem{PV_with_Batteries}
Y.~{Riffonneau}, S.~{Bacha}, F.~{Barruel}, and S.~{Ploix}, ``Optimal power flow
  management for grid connected pv systems with batteries,'' {\em IEEE
  Transactions on Sustainable Energy}, vol.~2, no.~3, pp.~309--320, 2011.

\bibitem{DC_microgrid}
T.~{Morstyn}, B.~{Hredzak}, and V.~G. {Agelidis}, ``Dynamic optimal power flow
  for dc microgrids with distributed battery energy storage systems,'' in {\em
  2016 IEEE Energy Conversion Congress and Exposition (ECCE)}, pp.~1--6, 2016.

\bibitem{OPF_in_MG}
Y.~{Levron}, J.~M. {Guerrero}, and Y.~{Beck}, ``Optimal power flow in
  microgrids with energy storage,'' {\em IEEE Transactions on Power Systems},
  vol.~28, no.~3, pp.~3226--3234, 2013.

\bibitem{nguyen2012new}
M.~Y. Nguyen, D.~H. Nguyen, and Y.~T. Yoon, ``A new battery energy storage
  charging/discharging scheme for wind power producers in real-time markets,''
  {\em Energies}, vol.~5, no.~12, pp.~5439--5452, 2012.

\bibitem{Optimal_control_rw}
D.~{Rosewater}, A.~{Headley}, F.~A. {Mier}, and S.~{Santoso}, ``Optimal control
  of a battery energy storage system with a charge-temperature-health model,''
  in {\em 2019 IEEE Power Energy Society General Meeting (PESGM)}, pp.~1--5,
  2019.

\bibitem{hector}
H.~E. {Perez}, X.~{Hu}, S.~{Dey}, and S.~J. {Moura}, ``Optimal charging of
  li-ion batteries with coupled electro-thermal-aging dynamics,'' {\em IEEE
  Transactions on Vehicular Technology}, vol.~66, no.~9, pp.~7761--7770, 2017.

\bibitem{jasinski}
S.~A. Jasinski, ``Modeling temperature distribution in cylindrical lithium ion
  batteries for use in electric vehicle cooling system design,'' 2008.

\bibitem{rosewater}
D.~M. {Rosewater}, D.~A. {Copp}, T.~A. {Nguyen}, R.~H. {Byrne}, and
  S.~{Santoso}, ``Battery energy storage models for optimal control,'' {\em
  IEEE Access}, vol.~7, pp.~178357--178391, 2019.

\bibitem{nguyen2018constructing}
H.~D. Nguyen, K.~Dvijotham, and K.~Turitsyn, ``Constructing convex inner
  approximations of steady-state security regions,'' {\em IEEE Trans. on Power
  Systems}, vol.~34, no.~1, pp.~257--267, 2018.

\bibitem{HungGPOPF}
P.~{Pareek} and H.~{D. Nguyen}, ``Gaussian process learning-based probabilistic
  optimal power flow,'' {\em IEEE Trans. on Power Systems}, 2020.

\bibitem{baran1989}
M.~E. Baran and F.~F. Wu, ``Network reconfiguration in distribution systems for
  loss reduction and load balancing,'' {\em IEEE Power Engineering Review},
  vol.~9, no.~4, pp.~101--102, 1989.

\end{thebibliography}

\end{document}